\documentclass[letterpaper,twocolumn,aps]{revtex4}
\usepackage{graphicx}
\def\frontmatter@finalspace{\addvspace{1\p@}}
\begin{document}
\def\baselinestreatch{1.3}
\def\TN{{T_{nat}}}
\def\TD{{T_{drv}}}
\def\Tav{{T_{ave}}}
\def\etal{{\it et. al.}}
\def\Dpqn{{\Delta^{p,q}_n}}
\title{Dynamical Phase Transitions In Driven Integrate-And-Fire Neurons}
\author{Jan R.~Engelbrecht$^{1}$ and Renato Mirollo$^{2}$}
\address{$^{1}$Department of Physics, Boston College, Chestnut Hill, MA 02467}
\address{$^{2}$Department of Mathematics, Boston College, Chestnut Hill, MA 02467}
\begin{abstract}
We explore the dynamics 
of an integrate-and-fire neuron with an oscillatory stimulus.
The frustration due to the competition between the neuron's natural
firing period and that of the oscillatory rhythm, leads to a rich structure of
asymptotic phase locking patterns and ordering dynamics. 
The phase transitions between these states 
can be classified as either tangent or
discontinuous bifurcations,
each with its own characteristic scaling laws.
The discontinuous bifurcations exhibit a new kind of
phase transition that may be viewed as intermediate between continuous and
first order, while tangent bifurcations behave like continuous transitions
with a diverging coherence scale.
\end{abstract} 
\maketitle 

Neurons in awake, behaving mammals receive complicated dendritic input
currents and respond with highly irregular trains of action potentials.
Unraveling the meaning of each neuron's train of spikes is a formidable
challenge.  A simple starting point is to characterize a neuron's activity
in terms of its firing rate (which varies in time in behaving organisms),
or in terms of temporal correlations between its spike times and that of
other neurons, either individually or collectively in a local rhythm.
There is a long history of detecting rhythmic neural activity at various scales,
from electroencephalography (EEG), to local field potentials, to oscillatory
membrane currents stimulating individual pyramidal and interneurons in
voltage clamp recordings.

In this letter we present a biophysical approach to explore consequences
of rhythmic inputs on the rate and timing of a model neuron's spikes.
Our framework is directly relevant for describing the behavior
of a neuron in a slice preparation with blocked dendritic inputs
and controlled injection of a simple stimulus current in a whole-cell 
patch-clamp setting.  Our analysis may also shed light on the
time scales and (transient) dynamical patterns in the spike trains that
may develop for the more complicated stimuli neurons receive in vivo.
Advocating a statistical-mechanical perspective, where minimal models 
yield insight into universal behavior of more realistic models,
we consider the one-dimensional integrate-and-fire (IF) model\cite{abbott}.
An IF model neuron receiving a constant current stimulus has
a constant firing rate.  We consider the consequences of an
additional small oscillatory input;
this second, competing time scale 
introduces frustration that results in pattern formation.  We explore 
connections between these stable patterns and the description of
critical phenomena associated with continuous phase transitions.

The IF model describes the response of a cell's
membrane potential $v(t)$ to an influx of current $I$. 
The voltage evolves according to the differential equation
$
\tau(dv/dt)=-(v-v_{eq})+RI
$,
together with the condition that 
when $v(t)$ reaches a threshold $v_{th}$, 
an instantaneous action potential is generated and $v(t)$
is reset to an equilibrium resting potential $v_{eq}$.
We use units where $R$ is the membrane resistance and the 
time constant $\tau=RC$ is proportional to the membrane capacitance $C$.
The parameters $R$, $I$, $\tau$, $v_{eq}$ and $v_{th}$
are all time-independent.
We set $\Delta v=v_{th}-v_{eq}$
and assume throughout that $\Delta v>0$.
If we start with an initial condition $v(t_0)=v_{eq}$,
then the potential $v(t)$ will
always reach threshold provided $RI>\Delta v$.
In this case 
the neuron fires at a constant rate,  with interspike interval
$
\TN=-\tau\log(1-{\Delta v/RI}) 
\label{T}
.
$ 

Next we consider the effect of adding a secondary 
periodic stimulus current $E\cos\omega t$,
which introduces a competing time scale $\TD=2\pi/\omega$.
The governing equation is now
\begin{equation}
\tau{dv\over dt}=-(v-v_{eq})+RI+E\cos\omega t
\label{IFE}
.
\end{equation}
We view the rhythm $E\cos\omega t$ as a perturbation of the constant current
term, and in this spirit
limit our attention to the case $0\le E<RI$.
The introduction of a second, competing time scale leads to a loss of the
simple periodic behavior of the original model.
As we shall see, the model neuron typically no longer has a constant
interspike interval, and can exhibit both periodic and aperiodic firing
patterns.

We again start with $v(t_0)=v_{eq}$
(changing $t_0$ allows us in effect to adjust the relative phase of the cosine 
drive at the initial condition).
The next spike time $t_1$ is the first solution to
$v(t)=v_{th}$ with $t>t_0$;
a solution exists
provided $RI+E/\sqrt{(\omega\tau)^2+1}>\Delta v$.
The dependence of $t_1$ on $t_0$ determines
a return map $F$; iterating the map $F$ generates a
spike train $t_0<t_1<t_2\cdots$ via
$
t_n=F(t_{n-1})=F^n(t_0)
.
$

\begin{figure}[b]
\vspace{-16pt}
\includegraphics*[height=1.9in]{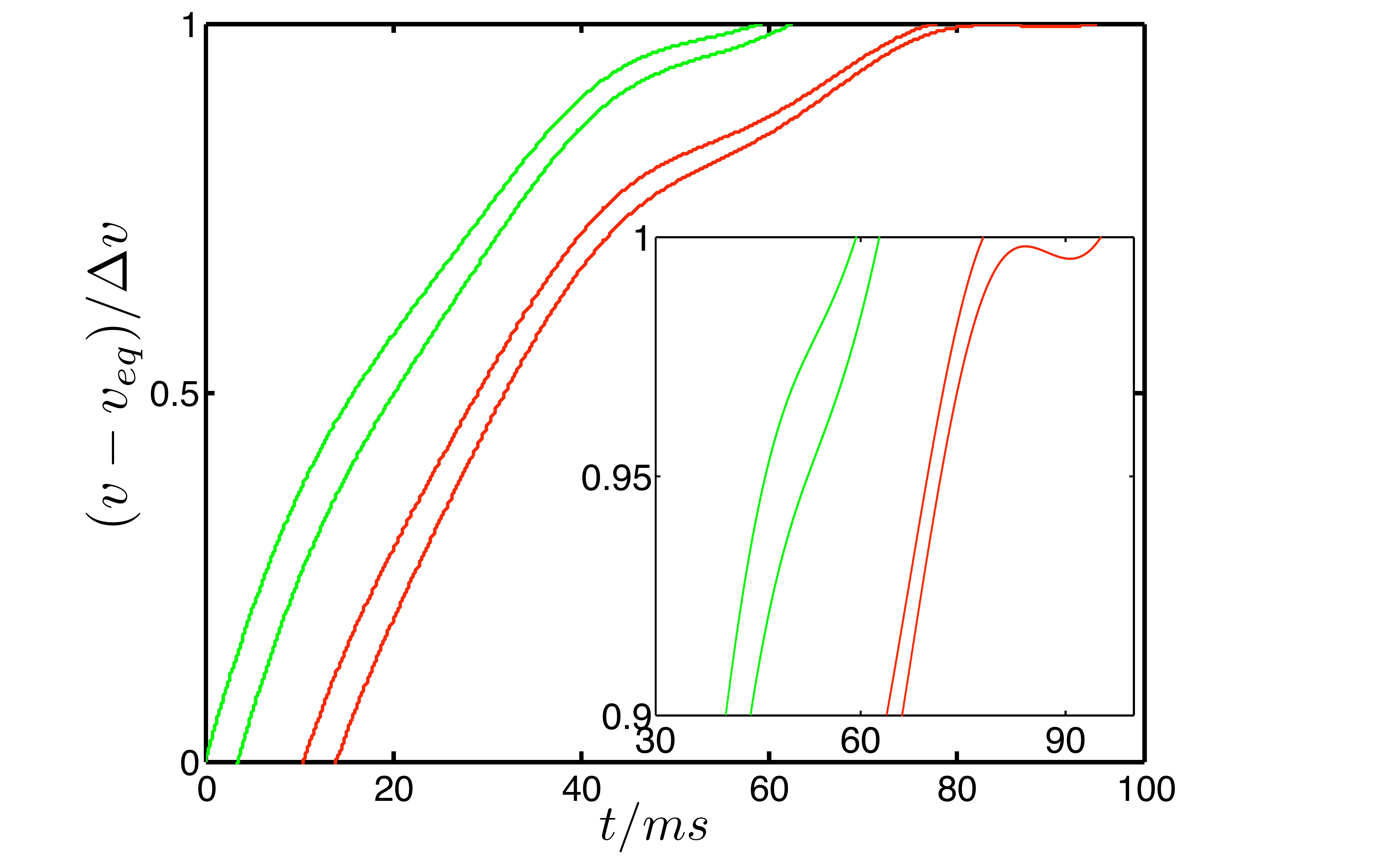}
\vspace{-14pt}
\caption{Solutions to (\ref{IFE}) when (\ref{condM})
holds (green) and fails (red).
\label{vt}
}
\vspace{-12pt}
\end{figure}

The map $F$ is continuous
when 
\begin{equation}
RI \ge E+\Delta v
\label{condM}
\end{equation}
(this implies $\dot v>0$ for $v<v_{th}$)
and is discontinuous otherwise.
The source of this discontinuity is illustrated in
Fig.~\ref{vt}, which shows solutions to 
(\ref{IFE})
with nearly identical initial conditions in the two 
different cases where condition
(\ref{condM}) is satisfied (green) and not satisfied (red).
When (\ref{condM}) fails (\ref{IFE}) has a solution with
$\dot v=0$ at threshold, which causes a jump discontinuity in 
the return map $F$.

We are interested in the evolution of the interspike intervals (ISIs) 
$t_n-t_{n-1}$
for $E>0$, especially whether these intervals become periodic and if so,
how rapidly such a pattern is established.
The asymptotic dynamics are determined by
the average interspike interval
\begin{equation}
\Tav=\lim_{n\to\infty}{t_n\over n}
=\lim_{n\to\infty}{F^n(t_0)\over n}
\label{Tav}
\end{equation}
i.e., the inverse of the neuron's firing rate.

The map $F$ satisfies the periodicity relation 
$F(t+\TD)=F(t)+\TD$
reflecting the periodicity of the drive term.
Consequently, from the theory of circle maps~\cite{devaney,knight,keener0},
the limit defining $\Tav$ exists, is independent of the initial condition 
$t_0$ and depends continuously on the parameters 
($RI$, $\tau$, $\omega$ and $E$).
Furthermore, the dimensionless ratio $\Tav/\TD$ is a rational number $r=p/q$
iff 
\begin{equation}
F^q(t^*)=t^*+p\;\TD
\label{rat}
\end{equation}
for some $t^*$; in other words 
$t^*$ 
is a fixed point of the map $F^q(t)-p\;\TD$.
So the spike train beginning with $t_0=t^*$ satisfies
$t_{n+q}=t_n+p\TD$ and consequently the sequence of
{\it phases} of $t_n$ relative to $\TD$ repeats every $q$ firings.

\begin{figure}
\includegraphics*[height=2.3in]{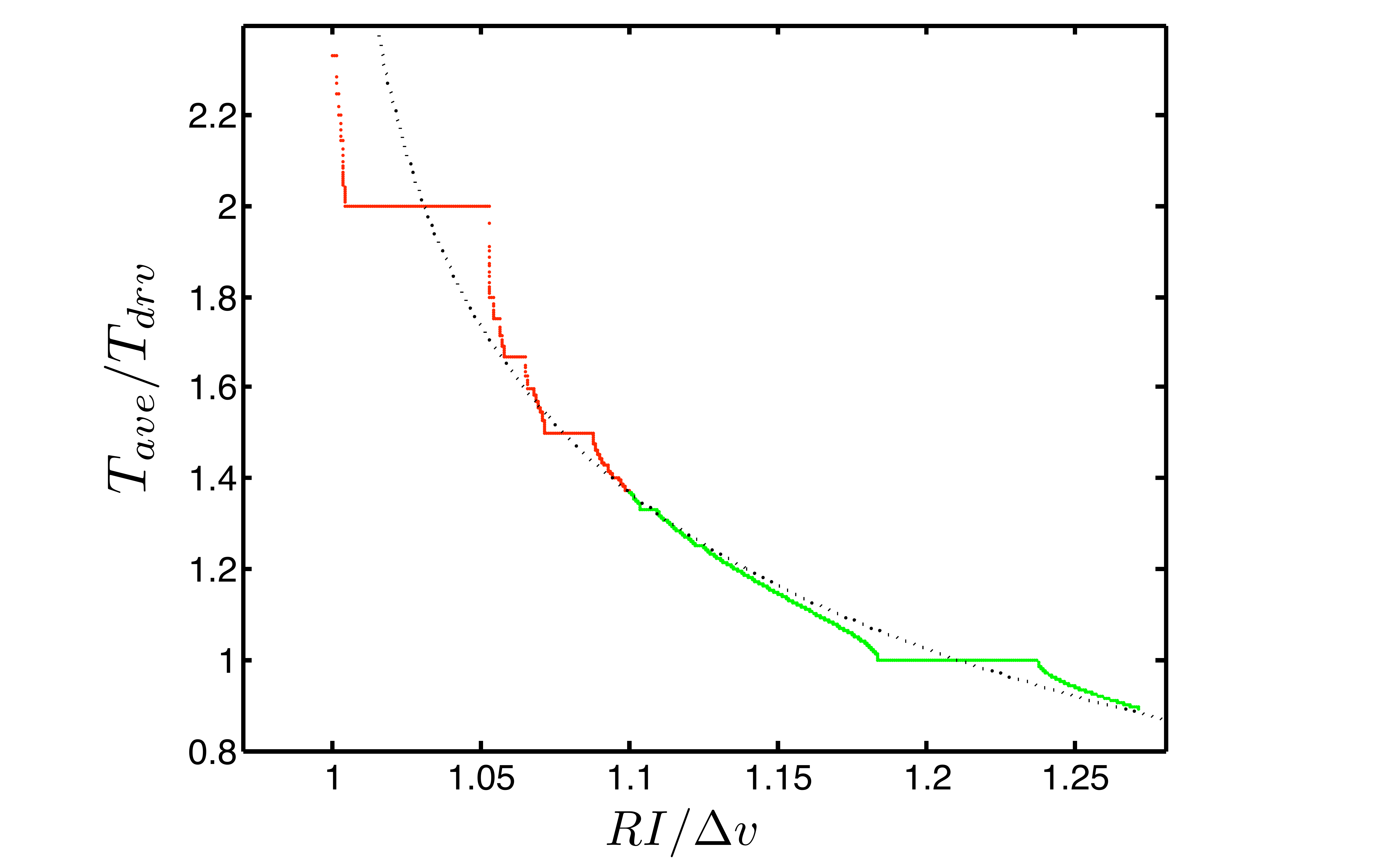}
\vspace{-12pt}
\caption{Average period in units of the drive period vs.~the parameter $RI$
for drive amplitude $E=0.1\Delta v$, $\tau=20ms$ and $\TD=35ms$.
The dotted curve is
$\TN/\TD$, which is proportional to the period in the undriven case.
\label{plateaux}
}
\vspace{-12pt}
\end{figure}

In Fig.~\ref{plateaux} we plot $\Tav/\TD$ as a function of
the parameter $RI$,
while keeping fixed $\TD=35\;ms$, $\tau=20\;ms$ and $E=0.1\;\Delta v$.
We divide the graph into two regions
according to whether the return map is continuous (green) or discontinuous (red).
For a given number $r$, let $RI_r^-$ and $RI_r^+$ 
denote the minimum and maximum values 
of $RI$ for which $\Tav/\TD=r$.
$RI_r^-<RI_r^+$ iff $r$ is a rational number; in other words
the plateaux in Fig.~\ref{plateaux} correspond to rational multiples of the 
drive period. 
When $r$ is an integer and the return map is continuous (green part),
$RI_r^{\pm}$ can be determined algebraically; in this special case
\begin{equation}
RI_r^{\pm}={\Delta v \over
1-e^{-2\pi r/\omega\tau} }
\pm{E\over\sqrt{(\omega\tau)^2+1}}
\label{Ic}
,
\end{equation}
and $\Tav=\TN$ at the midpoint of the plateau.
The width of such a plateau is then proportional to $E$, the magnitude of the 
oscillatory drive.
In all other cases, $RI_r^{\pm}$ needs to be determined numerically.

The asymptotic structure of the average firing rate as
in Fig.~\ref{plateaux} has been known for some 
time\cite{knight,keener0,keener}.
In this paper our focus is on {\it the approach} to the asymptotic behavior,
and the resulting connection to dynamical {\it phase transitions}.
The circle map theorem\cite{devaney} guarantees that the asymptotic value of
$\Tav$ for a given $RI$ 
is independent of the initial condition $t_0$.
This stability allows us to view  
Fig.~\ref{plateaux} as a phase diagram, with each plateau 
a state corresponding to some rational number $r$ which we express
as a fraction $r=p/q$ (in lowest terms).
These states are analogous to phases of matter and 
the boundaries of the plateaux to phase transitions.  

Within each $p/q$ entrainment plateau, 
the spike train converges to a periodic pattern of ISI's that repeats every $q$
spikes, corresponding to a stable fixed point of $F^q(t)-p\TD$.
For perfect $p/q$ entrainment
$t_{n+q}-t_n-p\TD=0$; hence 
\begin{equation}
\Dpqn=t_{n+q}-t_n-p\TD
\label{Spqn}
\end{equation}
measures the deviation from $p/q$ entrainment.
The convergence (within a plateau) is geometric, so
$\Dpqn \sim x^{-n}.$

Henceforth we fix
$\tau$, $\omega$ and $E$ and consider the parametric dependence on
$RI$.
The fixed points of
$F^q(t)-p\TD$ vary with $RI$ and ultimately vanish through 
some kind of bifurcation at the edges of the $p/q$ entrainment plateau.
Analogous to the theory of phase transitions, 
the equation
\begin{equation}
\Dpqn\sim 
e^{-n/\xi_\tau(RI)} 
\label{xi}
\end{equation}
defines a coherence time $\xi_\tau(RI)$
that characterizes how rapidly the phase-locked solution is approached.
Of particular interest is how $\xi_\tau(RI)$ scales with respect to
the tuning parameter $RI$ as an edge of an entrainment 
plateau (phase boundary) is approached.
As we shall show, 
the scaling has a universal form dictated by the type of bifurcation
through which the fixed points are lost.

\begin{figure}
\includegraphics*[height=1.9in]{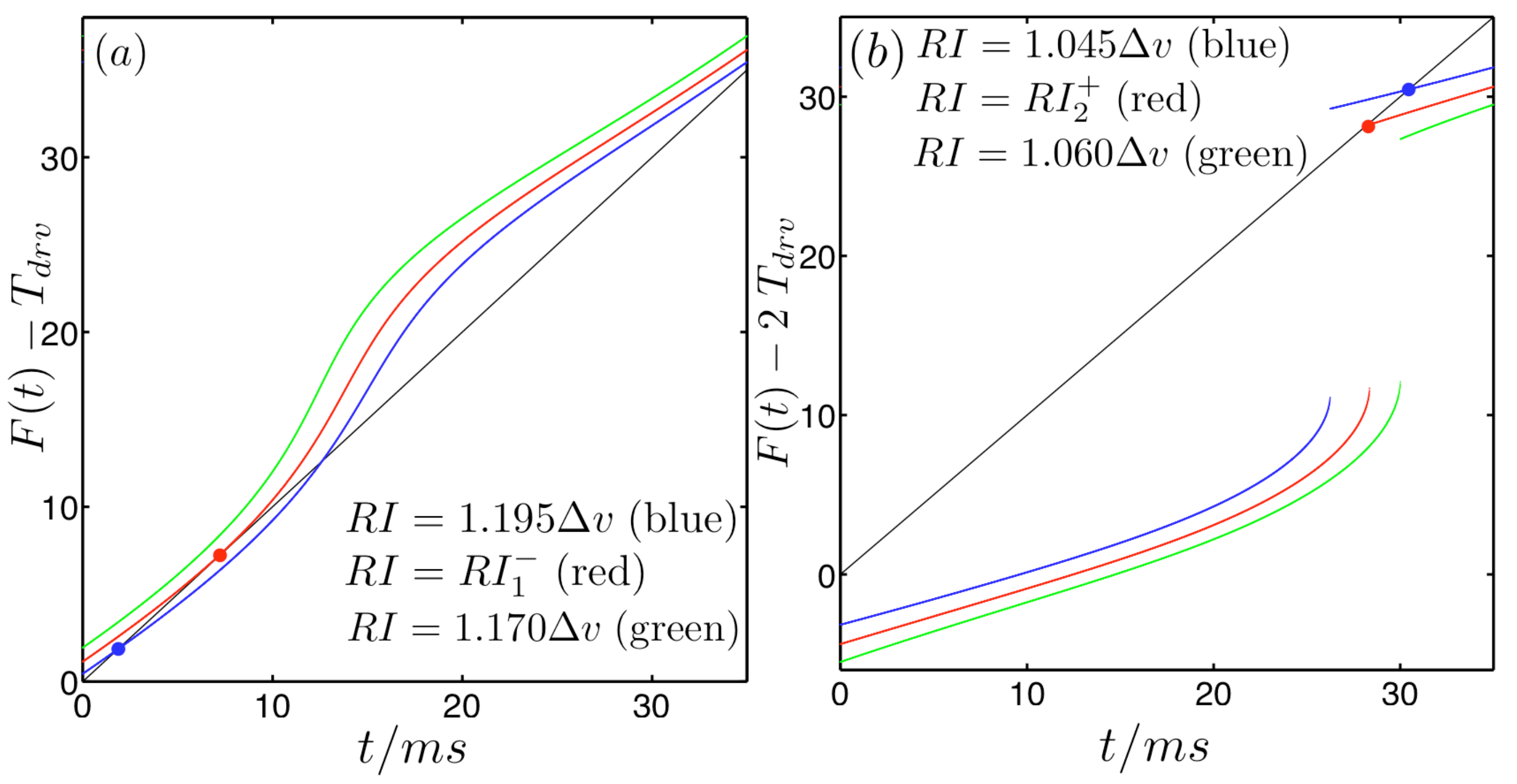}
\vspace{-16pt}
\caption{(a) Return maps associated with the tangent bifurcation at left edge 
of the $r=1$ plateau;
(b) Return maps associated with the discontinuous bifurcation at right edge 
of the $r=2$ plateau.
\label{maps}
}
\vspace{-12pt}
\end{figure}

We first analyze the phase transitions and associated scaling behaviors 
that occur in the parameter range when the return map $F$ is continuous,
as in Fig.~\ref{maps}(a).
Upon varying $RI$ the fixed point is here lost through a {\it tangent} 
bifurcation.
The generic behavior near such a phase boundary is modeled by
the simple map 
$g(x)=x-x^2+\lambda$, which has a stable fixed point at
$x^*=\sqrt{\lambda}$
that is lost through a tangent bifurcation as the control parameter 
$\lambda\to0$. Let $\delta x_n=x_n-x^*$; then 
$\delta x_{n+1}-\delta x_n=-2\sqrt{\lambda}\;\delta x_n-(\delta x_n)^2$,
which has large $n$ solution 
$\delta x_{n}\sim\exp(-2\sqrt{\lambda}\;n)$ 
so $x_n\to x^*$ with a coherence time
$\xi\sim1/\sqrt{\lambda}$.
This generic scaling holds for each fixed point of $F^q(t)-p\TD$ and
as the phase boundary is approached from within a plateau the coherence time 
then scales as
\begin{equation}
\xi_\tau(RI)\sim
{1\over |RI-RI_r^{\pm}|^{\frac12}}
\label{nu12}
\end{equation}
consistent with classical exponent $\nu=\frac12$ in equilibrium critical
phenomena\cite{Nigel}.

The coherence time diverges at
the phase boundary 
(when the control parameter $RI=RI_r^{\pm}$),
and the dynamics can again be modeled by the map $g(x)$.
In this case $\lambda=0$, the fixed point $x^*=0$
and the $\delta x_n^2$ term above becomes relevant.
The large $n$ solution 
is now $x_n\sim1/n$ and hence $x_{n+1}-x_n\sim-1/n^2$;
in particular the fixed point at the tangent bifurcation is no longer
approached geometrically. 
Analogously,
the $p/q$-entrainment coherence at the phase boundary
then develops according to the power law 
\begin{equation}
\Dpqn\sim 
{1\over n^{2}}
\label{eta0}
\end{equation}
consistent with a critical exponent $\eta=0$.

As we vary the control parameter $RI$ so as to
exit the $p/q$ entrainment plateau, 
``bottlenecks'' develop near the locations of the $q$ lost fixed points.
The resulting dynamics can again be modeled by
the map $g(x)$ which has a bottleneck near $x=0$
for small negative $\lambda$.
As $\lambda\to0^-$,
the number of iterations $N_\lambda$ needed to pass through
a fixed interval $[-c,c]$ around zero scales like 
$N_\lambda\sim1/\sqrt{|\lambda|}$.
$N_\lambda$ characterizes how long in takes to pass through a single bottleneck
and introduces a time-scale outside the entrainment plateau that diverges
similar to the coherence time in eqn.~(\ref{nu12}).

\begin{figure}
\includegraphics*[height=2.0in]{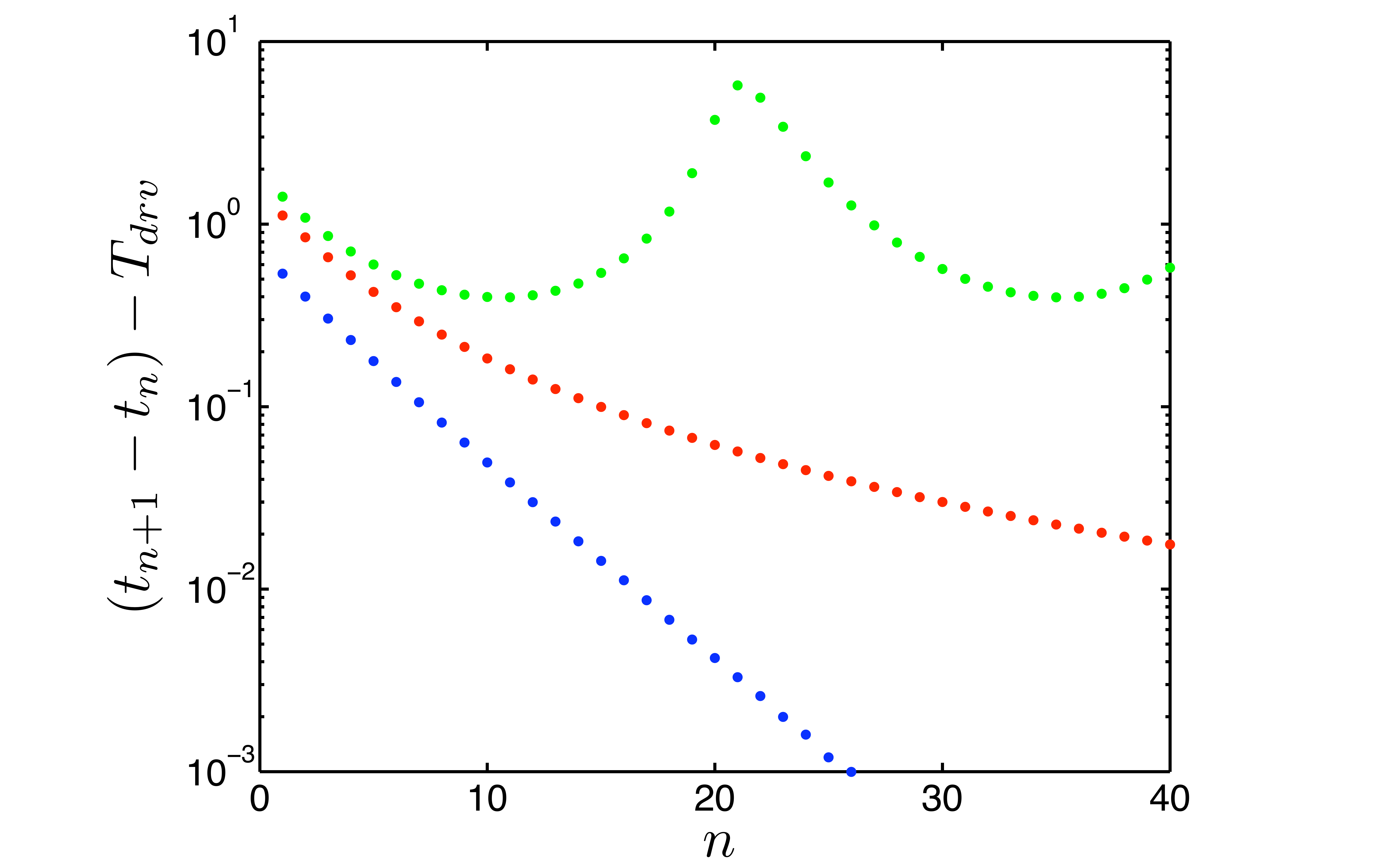}
\vspace{-12pt}
\caption{Approach to entrainment in ISI's 
($\Delta_n^{1,1}$) inside (green), at (blue) and outside (red)
the left edge of the 1/1 plateau, respectively exhibiting scaling of (\ref{xi}),
(\ref{eta0}) and bottleneck behavior.
\label{ISI}
}
\vspace{-12pt}
\end{figure}

The approach to 
$p/q$ entrainment 
is illustrated
in Fig.~\ref{ISI} which demonstrates exponentially fast coherence
of the form
(\ref{xi}) inside the 1-1-plateau, the power-law form (\ref{eta0}) at the
phase boundary and bottleneck behavior just outside the 
1-1-entrainment phase.
The repeated bottleneck behavior represents failed entrainment to
this particular $p/q$-entrainment phase.
If this pattern eventually repeats periodically, 
the average period will converge to different rational multiple of $\TD$
and hence lie on a different $p'/q'$-entrainment plateau;
otherwise $\Tav/\TD$ is irrational.

\begin{figure}
\includegraphics*[height=3.6in]{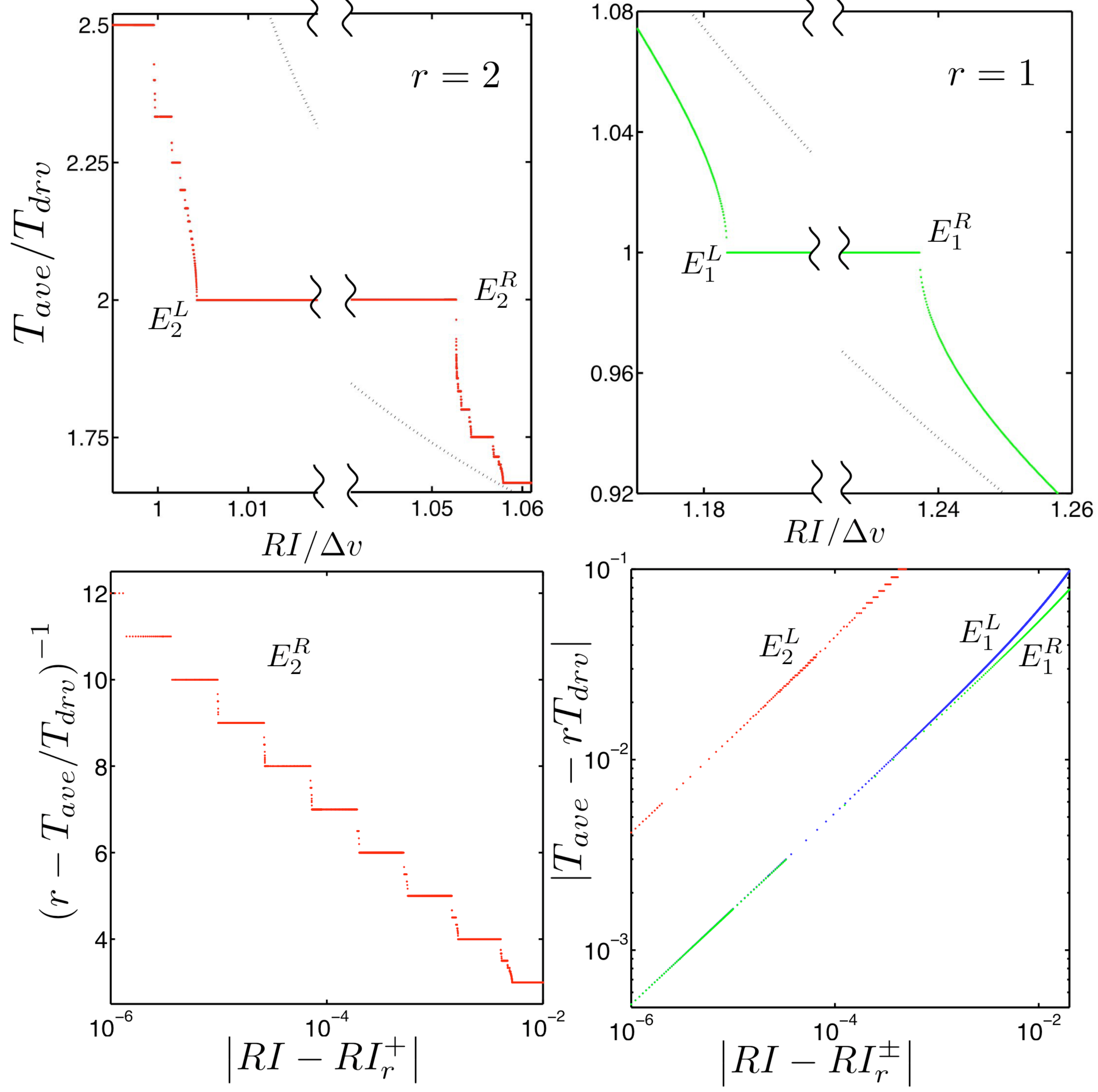}
\vspace{-20pt}
\caption{Scaling in $\Tav$ near the edges of the $r=2$ and $r=1$ plateaux.
Edge $E_2^R$ demonstrates the logarithmic scaling in (\ref{logscaling}) at a 
discontinuous bifurcation, while edges $E_2^L$, $E_1^L$ and $E_1^R$ exhibit 
power law scaling in (\ref{beta12}) at tangent bifurcations.
\label{opscaling}
}
\vspace{-12pt}
\end{figure}

For $RI$ just outside the plateau, the deviation of the average period from
$r\TD$ is inversely proportional to the number of iterations
required to pass through the bottleneck caused by lost fixed point
of $F^q(t)-p\TD$; thus
\begin{equation}
\left|\Tav-{r\TD}\right|
\sim
{|RI-RI_r^{\pm}|^{\frac12}}
\label{beta12}
.
\end{equation}
This average deviation of entrainment plays the role of a
disorder parameter which, in analogy to equilibrium phase transitions,
identifies the exponent $\beta=\frac12$.
This scaling form is illustrated in Fig.~\ref{opscaling}.

Generically, a smooth map $F$
can acquire or lose periodic points
only through
tangent bifurcations,
which dictate the universal scaling near the phase boundaries described above.
However, when the map is discontinuous 
it can also acquire or lose periodic points
through a {\it discontinuous} bifurcation.
As we shall see, the associated
scaling near the phase boundaries for these bifurcations 
belongs to a new universality class.

An example of this type of bifurcation is illustrated in 
Fig.~\ref{maps}(b), for $r=2$.
The behavior at the bifurcation (phase boundary) depicted here
differs fundamentally from 
the continuous case in Fig.~\ref{maps}(a), in that
the slope of the map at the fixed point here remains strictly less than one. 
Consequently, iterates of the map still
converge to the fixed point following the geometric form in (\ref{xi}),
corresponding to a finite coherence time at the bifurcation, as opposed to
power law scaling.  

To explore the behavior near the plateau edges for discontinuous bifurcations,
we introduce a simple map $h$ of the form $h(x)=ax+\lambda$ for $x>0$ (with
$0<a<1$) and
with a discontinuous jump at $x=0$.
As $\lambda$ sweeps through zero, the fixed point of $h(x)$ is
lost in a bifurcation similar to that seen in Fig.~\ref{maps}(b).
Since the map is linear for $x>0$, successive iterations satisfy
$x_n=h^n(x_0)=x^*+a^n(x_0-x^*)$ while $x_n>0$, where
$x^*=\lambda/(1-a)$ which is a fixed point of $h(x)$ for $\lambda>0$.
A bottleneck near $x=0$ again develops for $\lambda$ small negative, 
and even though $x^*<0$ is then not a fixed point of $h(x)$, 
it still controls the
passage through the bottleneck in terms of the expression for $x_n$ above.
The number of iterations $N_\lambda$ needed to pass through an 
interval $[0,c]$ is determined by solving
$0=h^n(c)$ for $n$ (and rounding up to the nearest integer).
As $\lambda\to0^-$, the solution
scales like $N_\lambda\sim-\ln|\lambda|$.

Consequently, for $RI$ just outside the edge of a plateau 
at which a discontinuous bifurcation occurs, 
the deviation of the average period from
$r\TD$ scales as
\begin{equation}
\left|\Tav-{r\TD}\right|
\sim
{-1\over
\ln |RI-RI_r^{\pm}|}
\label{logscaling}
.
\end{equation}
So our
disorder parameter vanishes {\it logarithmically} at phase boundaries 
determined by discontinuous bifurcations,
slower than any power law.

Note that the functions $F$ graphed in Fig.~\ref{maps}(b) are increasing and 
have
slope $+\infty$ on the left at the discontinuities.
Furthermore, the dependence of $F$ on the parameter $RI$ is such that
as $RI$ increases, the graph of $F$ moves downward.
These properties hold in general for all maps $F$ maps under consideration
as well as the iterates $F^q$.
At a bifurcation the map $F^q(t)-p\TD$ still
has fixed points but
must also lie completely on
one side of the line $y=t$.
The bifurcation is discontinuous if these fixed points occur at the
discontinuities of the map.
Since $F^q$ has slope $+\infty$ on the left at the jump discontinuities,
the fixed points at a discontinuous bifurcation must occur on the right side
of the jump discontinuities, and
the map $F^q(t)-p\TD$ must lie {\it below} the line $y=t$.
These fixed points are lost upon increasing $RI$ and
consequently discontinuous bifurcations can only occur on the right edges
of the plateaux, as is the case in the particular example 
illustrated above for $r=2$.
In other words, the bifurcations occurring at the
left edges of the entrainment plateaux are all
tangent bifurcations, even in the parameter range where $F$
is discontinuous.

On the other hand, both types of bifurcations occur 
at the right edges in the parameter range where $F$
is discontinuous, although tangent bifurcations are quite rare.
For example, in Fig.~\ref{plateaux} tangent bifurcations occur 
at the right edges of the plateaux 
for $r=\frac32, \frac75$ and $\frac{11}8$; all the other 
right-edge bifurcations we investigated in this
parameter range are discontinuous.
Furthermore, in this example, 
the bifurcations at the right edges are {\it all} discontinuous
for $RI$ below some threshold.
In fact, we can show that such a threshold exists
if the oscillatory drive $E$ is sufficiently small relative to $\Delta v$
(technically when
$E<\Delta v/(1+((\omega\tau)^2+1)^{-\frac12})$).
The point is
that under this condition,
the map $F$ is concave up everywhere for $RI$ sufficiently small
and since $F$ is increasing, the same holds for 
all its iterates.  This rules out the possibility of tangent bifurcations 
at right edges of plateaux; i.e.,
where the map $F^q(t)-p\TD$ is below the line $y=t$.

In conclusion, periodically driven IF neurons lose $p$-$q$ entrainment through
two markedly different routes, corresponding to the tangent and discontinuous
bifurcations described above.
Each has its own characteristic universal scaling laws which
measure the rate of convergence to entrainment within the $p$-$q$ plateau
as well as the deviation from $p$-$q$ entrainment just outside the plateau.

In equilibrium statistical mechanics,
a phase transition is classified as either 
{\it continuous}, which has a diverging coherence scale ($\xi$)
and a vanishing order 
parameter at its critical point, or
{\it first order}, which has no diverging coherence 
scale and a discontinuous
jump in the order parameter at its transition.
The scaling laws at tangent bifurcations are identical to that of a
continuous phase transition with `classical' exponents $\beta=\frac12$,
$\nu=\frac12$ and $\eta=0$.
However, the behavior at the discontinuous bifurcations does not match 
our conventional understanding of phase transitions with universal scaling.
The finite coherence time at the discontinuous bifurcations
is a feature of first order phase transitions, which have discontinuous 
jumps in their order parameters and no universal scaling laws.
But as we have seen, $\Tav$ varies 
continuously and hence the disorder parameter $\Tav-r\TD$ 
vanishes at the bifurcation, as is the case for continuous phase transitions.
The logarithmic scaling law for $\Tav-r\TD$ vanishes more slowly than any 
power law, and hence exhibits behavior which is intermediate
between conventional continuous and first order phase transitions.

For conventional continuous phase transitions, universal scaling results from
the singularity associated with a diverging coherence scale.  Our
discontinuous bifurcation does not have such a diverging scale, yet exhibits a
new kind of universal scaling.  Here it is the singularity in the
discontinuous map that is responsible for universality.

From a neuro-physiological perspective, our results suggest that when a neuron
firing with some rate receives an additional rhythmic stimulus, a range of
$p$-$q$ entrainment possibilities exist.  Moreover, the convergence to
entrainment to a $p$-$q$ phase-locked firing pattern is characterized by a
coherence time which depends sensitively on the distance to the entrainment
plateau edge.  Whole-cell slice recording, where an individual cell is
stimulated with a constant plus oscillatory current injection, would be an
ideal setting in which to explore in vitro the scaling and pattern formation
discussed in this letter.

JRE acknowledges very useful conversations with
John Hopfield and David Sherrington and support from ICAM.

\end{document}